# Spectral Precoding for Out-of-band Power Reduction under Condition Number Constraint in OFDM-Based System

Lebing Pan

[1] No.50 Research Institute of China Electronic Technology Group Corporation, Shanghai 200311, China.

forza@aliyun.com

**Abstract:** Due to the flexibility in spectrum shaping, orthogonal frequency division multiplexing (OFDM) is a promising technique for dynamic spectrum access. However, the out-of-band (OOB) power radiation of OFDM introduces significant interference to the adjacent users. This problem is serious in cognitive radio (CR) networks, which enables the secondary system to access the instantaneous spectrum hole. Existing methods either do not effectively reduce the OOB power leakage or introduce significant bit-error-rate (BER) performance deterioration in the receiver. In this paper, a joint spectral precoding (JSP) scheme is developed for OOB power reduction by the matrix operations of orthogonal projection and singular value decomposition (SVD). We also propose an algorithm to design the precoding matrix under receive performance constraint, which is converted to matrix condition number constraint in practice. This method well achieves the desirable spectrum envelope and receive performance by selecting zero-forcing frequencies. Simulation results show that the OOB power decreases significantly by the proposed scheme under condition number constraint.

## 1. Introduction

Dynamic spectrum access [1, 2] technology is extensively studied as an effective scheme to achieve high spectral efficiency, which is a crucial step for cognitive radio (CR) networks. Due to the flexible operability over non-continuous bands, orthogonal frequency division multiplexing (OFDM) is considered as a candidate transmission technology for CR system [3]. However, due to the use of rectangular pulse shaping, the power attenuation of its sidelobe is slow by the square of the distance to the main lobe in the frequency domain. Therefore, the out-of-band (OOB) power radiation or sidelobe leakage of OFDM causes severe interference to the adjacent users. Furthermore, this problem is serious in CR networks, which enables the secondary users to access the instantaneous spectrum hole. Therefore, these secondary users need to ensure that the interference level of the power emission is acceptable for primary users.

In practice, typically of the order of 10% guard-band is needed for an OFDM signal in the long term evolution (LTE) system [4]. Therefore, the spectrum efficiency is significantly reduced. The traditional method of sidelobe suppression is based on windowing techniques [5], such as the raised cosine windowing [6], which is applied to the time-domain signal wave. However, this scheme requires an extended guard interval to avoid signal distortion, and the spectrum efficiency is also reduced for large guard intervals. The cancellation carriers (CCs) [7, 8] technique inserts a few carriers at the edge of the spectrum in order to cancel the sidelobe of the data carriers. However, this technique degraded the signal-to-noise ratio (SNR) at the receiver. The subcarrier weighting method [9] is based on weighting the individual subcarriers in a way that the sidelobe of the transmission signal are minimized according to an optimization algorithm. However, the bit-error-rate (BER) increases in the receiver, and when the number of subcarriers is large, it is difficult to implement in real-time scenario. The multiple choice sequence method [10] maps the transmitted symbol into multiple equivalent transmit sequences. Therefore, the system throughput is reduced when the size of sequences set grows. Constellation adjustment [11] and constellation expansion [12] are difficult to implement when the order of quadrature amplitude modulation (QAM) is high. Strikingly, the methods [10, 11] require the transmission of side information. The adaptive symbol transition [13] scheme usually provides weak sidelobe suppression in frequency ranges closely neighboring the secondary user occupied band. However, the schemes [7-14] depend heavily on the transmitted data symbols.

The precoding technology is widely used in OFDM system to enhance the performance of transmission reliability over the wireless environment, in which there are many precoding methods proposed for OOB power reduction [15]. There are two main optimization schemes with obtaining large suppression performance. One is to force the frequency response of some frequency points to be zero by orthogonal projection [16-19]. These frequency points are regarded as zero-forcing frequencies. The other method [20, 21] is to minimize the power leakage in an optimization frequency region by adopting the quadratic optimization method, using matrix singular value decomposition (SVD). The precoding scheme in [16] is designed to satisfy the condition that the first $N$ derivatives of the signals are continuous at the edges of symbols. However, this method introduces an error floor in error performance. The sidelobe suppression with orthogonal projection (SSOP) method in [19] adopts one reserved subcarrier for recovering the distorted signal in the receiver. To maintain the BER performance, data cost is introduced in [17] by exploiting the redundant information in the subsequent OFDM symbol. The sidelobe suppression in [21] is based on minimizing the OOB power by selecting some frequencies in an optimized region. The suppression problem is first treated as a matrix Frobenius norm minimization problem, and the optimal orthogonal precoding matrix is designed based on matrix SVD. In [20], an approach is proposed for multiuser cognitive radio system. This method ensures user independence by constructing individual precoder to render selected spectrum nulls. Unlike the methods that focus on minimizing or forcing the sidelobe to zero, the mask compliant precoder in [22] forces the spectrum below the mask by solving an optimization problem. However, the algorithm leads to high complexity.

In this paper, a spectral precoding scheme is proposed with matrix orthogonal projection and SVD for OOB reduction in OFDM-based system. The main idea of this scheme is to reduce the OOB power under the receive quality. The condition number of the precoding matrix indicates the BER loss in the receiver. Therefore, we develop an iteration algorithm to design the precoding matrix under matrix condition number constraint. The proposed method has an appropriate balance among suppression performance, spectral efficiency and receive quality. Consequently, it is flexible for practical implementation.

The rest of the paper is organized as follows. In Section 2, we introduce the system model of OOB power reduction by the spectral precoding method. In Section 3, we present the proposed spectral precoding approach. Next, in Section 4, we provide an iteration algorithm to design the precoding matrix according to the desirable spectrum envelope, spectral efficiency and BER performance. Simulations are presented in Section 5 to demonstrate the performance of the proposed method, followed by a summary in Section 6.

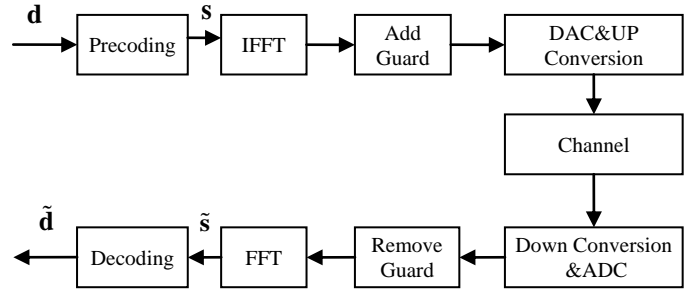

*Fig. 1 System diagram of spectrally precoded OFDM.*

## 2. System Model

The block diagram of a typical OFDM system using a precoding technique is illustrated in Fig. 1. The number of total carriers used in the transmitter is $M$. The digital spectral precoding process before inverse fast Fourier transform (IFFT) operation is expressed as

$$\mathbf{s} = \mathbf{Pd}, \quad (1)$$

where $\mathbf{d}$ is the original OFDM symbol of size $N \times 1$, and $\mathbf{s}$ is the precoded vector. The size of the precoding matrix $\mathbf{P}$ is $M \times N$ ($M \geq N$) and the coding redundancy $R = M - N$ is usually small. The spectral precoding method achieves better suppression performance in zero-padding (ZP) OFDM system than cyclic-prefix (CP) OFDM [19, 21]. What's more, the ZP scheme has already been proposed as an alternative to the CP in OFDM transmissions [23] and particularly for cognitive radio [24]. The proposed method also can be directly applied to CP systems with degraded performance on sidelobe suppression.

Conventionally, power spectral density (PSD) analysis for multicarrier systems is based on an analog model with a sinc kernel function [25]. The PSD converges to the sinc function with the sampling rate increasing. In addition, the oversampling constraint presented in [26] ensures the desirable power spectral sidelobe envelope property after precoding for DFT-based OFDM. In a general OFDM system, the time-domain signal can be defined by a rectangular function (baseband-equivalent) as

$$g(t) = \begin{cases} 1, & 0 \leq t \leq T \\ 0, & elsewhere \end{cases} \quad (2)$$

where $T$ is the symbol duration. The frequency domain representation of $m$-th subcarrier is written as

$$G_m(\omega) = e^{-j\omega T/2} \frac{\sin((\omega - \omega_m)T/2)}{(\omega - \omega_m)/2}, \quad (3)$$

where $\omega_m$ is the center frequency of $m$-th subcarrier. Therefore, the magnitude envelope in the OOB region ($\omega \neq \omega_m$) is expressed by

$$\begin{aligned} |G_m(\omega)| &= \left| \frac{\sin((\omega - \omega_m)T/2)}{(\omega - \omega_m)/2} \right| \\ &\leq \frac{2}{|\omega - \omega_m|}. \end{aligned} \quad (4)$$

Then we define the function $C_m(\omega)$ to indicate the magnitude envelope by

$$C_m(\omega) = \frac{1}{|\omega - \omega_m|}. \quad (5)$$

The expression (5) is obtained from (4) by ignoring a constant factor. This operation does not affect the mathematical analysis for the design of the precoding matrix [19]. The complex computing by (3) is converted to real operation, which reduce the complexity of spectral precoding in the following. Based on (5), using the superposition of $M$ carriers, the target function indicating the PSD at OOB frequency $\omega$ is defined by

Submitted

$$A(\omega) = \left|\sum_{m=0}^{M-1} s_m C_m(\omega)\right|^2, \quad (6)$$

where $s_m$ is the $m$-th element in the precoded OFDM vector $\mathbf{s}$. In OFDM system, the power spectrum of its sidelobe decays slowly as $(\Delta\omega)^{-2}$, where $\Delta\omega = \omega - \omega_m$ is the frequency distance to the mainlobe. From (1) and (6), the PSD target function of the OFDM signal is expressed in the matrix form as

$$P(\omega) = \frac{1}{T_s} E\{A(\omega)\} = \frac{1}{T_s} E\left\{\left|\mathbf{c}^T \mathbf{P}\mathbf{d}\right|^2\right\}, \quad (7)$$

where $\mathbf{c} = [C_0(\omega), C_1(\omega), ... C_{M-1}(\omega)]^T$, $(.)^T$ and $E$ denote transpose operation and expectation respectively. $P(\omega)$ and $C(\omega)$ are not the PSD presentation of the OFDM signal and the frequency response respectively, but indicate their envelope character. The goal of the precoding method is to design $\mathbf{P}$ to reduce the emission in OOB region. Simultaneously, the process is irrespective of the value of vector $\mathbf{d}$.

## 3. The Proposed Joint Spectral Precoding (JSP) Method

In this paper, a joint spectral precoding (JSP) method is proposed with two times orthogonal projection and one time SVD. The process is given by three steps: Inner orthogonal projection → SVD → Outside orthogonal projection. As presented in the previous papers [19, 21], in each step, the main idea of orthogonal projection is to force the power of zero-forcing frequency points to be zero, and the SVD operation is to minimize the power in optimized region. However, the precoding matrix derived from the final step does not achieve the goal in each step again. Unlike the methods that focus on minimizing or forcing the sidelobe to zero, we reduce the OOB power under the receive quality constraint by selecting the frequency points in each step. Therefore, the proposed method has an appropriate balance between the suppression performance and the receive quality. In addition, when the number of the reserved carriers is small, such as only one, then the number of zero-forcing frequencies in orthogonal projection method is one. But in the JSP, we set the zero-forcing frequency in inner and outside orthogonal projection to be different and better suppression performance is obtained.

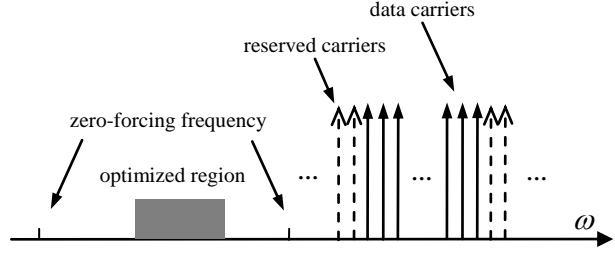

*Fig. 2 The spectrum diagram for OOB power reduction.*

Generally, the subcarriers of OFDM-based system are considered to be continuous. The proposed method also can be used for non-continuous multi-carriers system. The number of employed carriers is $M$ and the index is from $0$ to $M-1$. Fig. 2 illustrates the frequency region of $\omega < \omega_m$. The $R$ reserved carriers at the upper and lower spectral edges, which are used to achieve sidelobe suppression and maintain the receive quality.

Two groups of zero-forcing frequency points $\boldsymbol{\omega}_a$ and $\boldsymbol{\omega}_b$ are selected in inner and outside orthogonal projection respectively. The corresponding number is $N_a$ and $N_b$ ($N_a \leq R$, $N_b \leq R$). $K$ frequency points in the optimized region are chosen to reduce the OOB power by SVD operation. We first give the process of JSP in this section, and the detail of how to select the parameters is presented in the next.

In the first step, the group of zero-forcing frequency points $\boldsymbol{\omega}_a$ ($\boldsymbol{\omega}_a = [\omega_{a_1},...\omega_{a_i},...\omega_{a_{N_a}}]$, $\omega_{a_i} < \omega_0$ or $\omega_{a_i} > \omega_{M-1}$) are chosen to achieve

$$\mathbf{C}_a \mathbf{P}_a \hat{\mathbf{d}} = \mathbf{0}, \qquad (8)$$

Where $\hat{\mathbf{d}}$ is a vector of size $M$, derived from the original data symbol $\mathbf{d}_{N\times 1}$ by adding zero in the location of the reserved carriers, i.e., $\hat{\mathbf{d}} = [0,0,...\mathbf{d}^T,...0,0]^T$. $\mathbf{C}_a$ is the magnitude response matrix of size $N_a \times M$, whose element $C_a(i,j) = C_j(\omega_{a_i})$ computed by (5). $j$ is the index of the carriers. The solution of (8) is equivalent to map $\hat{\mathbf{d}}$ to the nullspace of $\mathbf{C}_a$, In [27], the orthogonal projector $\mathbf{P}_a$ mapping vector onto $\mathbf{C}_a^{\perp}$ ($\mathbf{C}_a^{\perp}$ is the nullspace of $\mathbf{C}_a$) along $\mathbf{C}_a$ is given by

Submitted$$\mathbf{P}_a = \mathbf{I}_M - \mathbf{C}_a(\mathbf{C}_a^T \mathbf{C}_a)^{-1} \mathbf{C}_a^T, \quad (9)$$

where $\mathbf{I}_M$ is a unit matrix of size $M \times M$. This projector has a property that the precoded data vector $\mathbf{P}_a \hat{\mathbf{d}}$, in the nullspace of $\mathbf{C}_a$, is closest to $\hat{\mathbf{d}}$ as

$$\min \|\mathbf{P}_a \hat{\mathbf{d}} - \hat{\mathbf{d}}\|. \qquad \mathbf{P}_a \hat{\mathbf{d}} \in \mathbf{C}_a^{\perp} \quad (10)$$

Obviously, $\mathbf{P}_a$ is non-orthogonal. Therefore, it will cause significant BER performance degradation in the receiver. After the step of the inner orthogonal projection, the precoded data is given by

$$\mathbf{s}_a = \mathbf{P}_a \hat{\mathbf{d}}. \quad (11)$$

Then in the second step, the magnitude response matrix of $K$ frequency points $\boldsymbol{\omega}_o$ in the optimized region is written as $\mathbf{C}_{op} = \mathbf{C}_o \mathbf{P}_a$, where $\mathbf{C}_o$ is a magnitude response matrix of the $K$ optimized frequency points. The elements in $\mathbf{C}_o$ is computed by (5). We then use the SVD operation to minimize the power leakage in the optimized region. The problem is determined by

$$\mathbf{P}_o = \arg\min_{\mathbf{P}} \|\mathbf{C}_{op}\mathbf{P}\|. \quad (12)$$

By decomposing the matrix $\mathbf{C}_{op}$ into $\mathbf{C}_{op} = \mathbf{U}_c \boldsymbol{\Sigma}_c \mathbf{V}_c^T$ using SVD, the optimal precoding matrix $\mathbf{P}_o$ to achieve (12) is derived from

$$\mathbf{P}_o = [\mathbf{V}_c^R, \mathbf{V}_c^{R+1}, ... \mathbf{V}_c^{M-1}], \quad (13)$$

where $\mathbf{V}_c^i$ is the $i$-th column of $\mathbf{V}_c$. The matrix $\mathbf{P}_o$ of size $M \times N$ is composed of the last $N$ columns in $\mathbf{V}_c$. $\mathbf{P}_o$ is an orthogonal matrix that $\mathbf{P}_o^T \mathbf{P}_o = \mathbf{I}_N$, where $\mathbf{I}_N$ is a unit matrix of size $N \times N$. The original data $\mathbf{d}$ is mapped to $\mathbf{P}_o \mathbf{d}$, while the length is extended from $N$ to $M$. The average power of each symbol of before and after precoding are $P_s \sum_{i=0}^{M-1} \rho_i^2$ and $P_s \sum_{i=0}^{N-1} \rho_i^2$, where $\rho_i$ is the $i$-th diagonal element in $\boldsymbol{\Sigma}_c$ and $P_s$ is the average power of each symbol $\mathbf{d}$. The second step is to abandon the largest $R$ singular values in $\mathbf{\Sigma}_c$, then the OOB power is reduced after precoding by $\mathbf{P}_o \mathbf{d}$.

In the third step, the group of zero-forcing frequency points $\mathbf{\omega}_b$ ($\mathbf{\omega}_b = [\omega_{b_1},...\omega_{b_i},...\omega_{b_{N_b}}]$, $\omega_{b_i} < \omega_0$ or $\omega_{b_i} > \omega_{M-1}$) are chosen. The orthogonal projection matrix $\mathbf{P}_b$ is obtained similar to (9) by

$$\mathbf{P}_b = \mathbf{I}_M - \mathbf{C}_b (\mathbf{C}_b^T \mathbf{C}_b)^{-1} \mathbf{C}_b^T, \qquad (14)$$

where the matrix $\mathbf{C}_b$ of size $N_b \times M$, is the magnitude response matrix of the frequency points $\omega_b$. Finally, the precoding matrix is obtained by

$$\mathbf{P} = \mathbf{P}_b \mathbf{P}_o. \qquad (15)$$

After precoding, the minimum error problem of decoding in the receiver, is given by

$$\min \| \hat{\mathbf{P}} \mathbf{P} \mathbf{d} - \mathbf{d} \|, \qquad (16)$$

where $\hat{\mathbf{P}}$ is the decoding matrix. The orthogonal projector $\mathbf{P}_b$ map the vector onto $\mathbf{C}_b^\perp$, so $rank(\mathbf{P}_b) = M - N_b \geq N$, where rank (.) denotes the rank of a matrix. Due to $rank(\mathbf{P}_o) = N$ and the value of $N_b$ is small, the case of that $\mathbf{P}$ is full column rank, is easy to be achieved in practice. Then the pseudo-inverse matrix of $\mathbf{P}$ is the optimal solution for (16) to achieve $\hat{\mathbf{P}} \mathbf{P} = \mathbf{I}_N$, which is given by $\hat{\mathbf{P}} = (\mathbf{P}^T \mathbf{P})^{-1} \mathbf{P}^T$.

This JSP scheme is unlike the methods that minimize the OOB power or forcing the sidelobe to zero. In the third step, we map $\mathbf{P}_o \mathbf{d}$ to the nullspace of $\mathbf{C}_b$ rather than the nullspace of $\mathbf{C}_b \mathbf{P}_o$. If the nullspace of $\mathbf{C}_b \mathbf{P}_o$ is selected, the JSP method turns to the traditional orthogonal projection method, which introduces large deterioration in the receiver. Furthermore, after the operations in the last two steps, the precoding matrix $\mathbf{P}$ does not achieve the goal in the first step of mapping $\mathbf{d}$ to the nullspace of $\mathbf{C}_a$. In addition, after the operation in the third step, $\mathbf{P}$ also does not achieve the goal of minimizing the power leakage in the optimized region in the second step.

As presented in the SVD operation [21] or orthogonal projection method [19], the receive quality and suppression performance have not been well balanced. We also had some tests to examine other combinations that only employing the first two steps or the last two steps in the JSP method. The results indicate that the suppression performance is similar to only using orthogonal projection or SVD operation. The reason is that the better OOB power reduction performance by the JSP method is obtained by two times orthogonal projection. In addition, the BER performance is improved by using reserved carriers in the second step. What's more, the desirable spectrum envelope also can be achieved by selecting the frequency points or optimized region in the three steps independently.

## 4. Design of The Precoding Matrix Under Condition Number Constraint

In this section, we develop an algorithm to design the precoding matrix, obtaining the desirable spectrum envelope under receive performance constraint. As illustrated in Fig.1, the data after FFT process in the receiver is expressed as

$$\tilde{\mathbf{s}} = \mathbf{Q}\mathbf{d} + \mathbf{n}, \quad (17)$$

where $\mathbf{Q} = \mathbf{HP}$, $\mathbf{H}$ is a complex diagonal matrix with the channel frequency response of $M$ subcarriers and $\mathbf{n}$ is complex additive white Gaussian noise (AWGN) vector with zero mean. $\mathbf{Q} = \mathbf{P}$ for AWGN channel. $\tilde{\mathbf{s}}$ is the received signal from noise measurement.

The matrix $\mathbf{P}$ is full column but $\mathbf{P}^T \mathbf{P} \neq \mathbf{I}_N$. Thus, this non-orthogonal precoding matrix will introduce BER loss in the receiver. If some singular values of $\mathbf{P}$ are too small, compared to the other values, low additive noise will result of large errors. Therefore, the condition number $\alpha$ of $\mathbf{P}$ is introduced, which is to measure the sensitivity of the solution of linear equations to errors in the data [28]. $\alpha$ is given by

$$\alpha = \text{Con}(\mathbf{P}) = \frac{\lambda_{\max}}{\lambda_{\min}}, \quad (18)$$

where $\lambda_{max}$ and $\lambda_{min}$ is the largest and the smallest singular value of $\mathbf{P}$. Con(.) denotes the condition number of a matrix. The value of $\alpha$ indicates the BER loss in the precoding. $\alpha \in [1, +\infty)$ and larger value of $\alpha$ leads to worse BER performance.

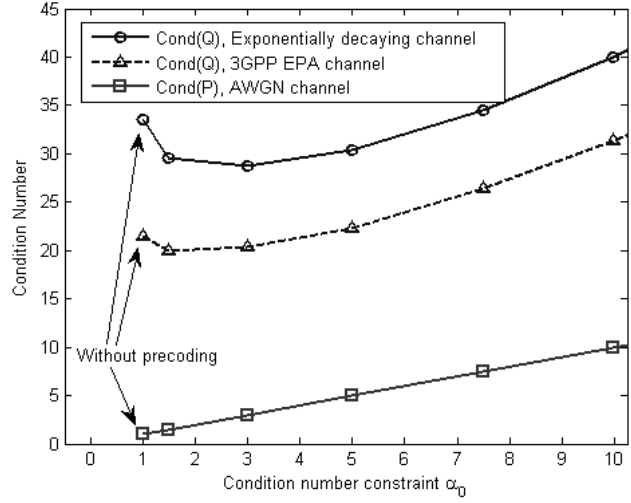

***Fig. 3** The condition number of transition matrix $\mathbf{Q}$ in AWGN channel and Rayleigh fading channel.*

The transition matrix $\mathbf{Q}$ significantly influence the receive quality. We examine the average condition number of $\mathbf{Q}$ through different channels in Fig. 3. The results of through Rayleigh channel are averaged over 2000 realizations. Two Rayleigh channel is selected: 3GPP extended pedestrian A (EPA) model [29], whose excess tap delay = (0 30 70 90 110 190 410)ns with the relative power = (0 -1 -2 -3 -8 -17.2 -20.8)dB, and a ten-tap Rayleigh block fading channel with exponentially decaying powers set as $E(|h_l(i)|^2) = e^{-l/3} / \sum_{j=0}^{9} e^{-j/3}$, $j = 0,1,...,9$. As illustrated in Fig. 3, the receive performance is better through the AWGN channel than the Rayleigh fading channel. Compared to the transmission without precoding, the condition number of $\mathbf{Q}$ linearly increases with condition number constraint $\alpha_0$ in AWGN channel, but decrease in the Rayleigh fading channel when the value of $\alpha_0$ is small. The zero-forcing equalizer is used for AWGN channel and fading channel respectively by

Submitted

$$\tilde{\mathbf{d}} = \begin{cases} (\mathbf{P}^T\mathbf{P})^{-1}\mathbf{P}^T\tilde{\mathbf{s}}, & AWGN\ channel \\ (\mathbf{Q}^H\mathbf{Q})^{-1}\mathbf{Q}^H\tilde{\mathbf{s}}, & Fading\ channel \end{cases} \quad (19)$$

where $(.)^H$ and $\tilde{\mathbf{d}}$ denotes conjugate transpose and the decoded data respectively.

In this part, we develop an algorithm to obtain $\mathbf{P}$ under a condition number constraint by selecting the frequency points. In order to keep the receive performance decreases slightly, the number of zero-forcing frequency points in the first orthogonal projection is chosen as small as possible. In practice, we select $N_a = 1$ for single-side suppression or $N_a = 2$ for double-side suppression to keep the spectrum symmetric. If the special case of $R = 1$ is selected, then we fix $N_a = 1$. The envelope of the precoded PSD curve is mainly determined by the outside orthogonal projection. Thus, $N_b = R$ is selected for high suppression performance. The two group zero-forcing frequencies $\boldsymbol{\omega}_a$ and $\boldsymbol{\omega}_b$ are arranged in the different OOB region, far from or close to the mainlobe. The one far from the mainlobe is fixed, for the amplitude response is weak by (5). The one closed to the mainlobe is used to maintain the receive quality by adjusting its location. In the SVD operation, in order to effectively abandon the largest $R$ singular values in $\boldsymbol{\Sigma}_c$, $K$ should larger or equal to $R$. We select $\boldsymbol{\omega}_o = \boldsymbol{\omega}_b$ for simple implementation and $K = R$. Then the variables for obtaining $\mathbf{P}$ are only the group closed to the mainlobe. Thus, we change these frequency points to achieve

$$\mathbf{P} = \max_{\mathbf{P}} \mathrm{Con}(\mathbf{P}) \quad s.t.\ \mathrm{Con}(\mathbf{P}) \leq \alpha_0, \quad (20)$$

where $\alpha_0$ is the given condition number according to the BER constraint. The signal estimation performance under condition number constraint in a communication system is illustrated in [28, 30]. In addition, as presented in [16, 19], the zero-forcing frequency point close to mainlobe leads to quicker power reduction on the edge of the mainlobe, but the power far from the mainlobe is larger. Inversely, if these points are far from the mainlobe, the power decreases slowly on the edge, while the emission from the mainlobe decreases largely.

**Table 1** The correlation between OFDM performance and the parameters in the proposed method.

|  | Case | $\alpha_0$ | $R$ | $N_a$ |
|---|---|---|---|---|
| BER | — | ↓ | — | ↓ |
| OOB power reduction | Case A: Quicker power reduction on the edge of the mainlobe<br>Case B: Lower power leakage far from the mainlobe | ↑ | ↑ | Na=1: Single-side suppression<br>Na=2: Double-side suppression |

↓ : Negative correlation;  ↑ : Positive correlation;  — : Weak or no effect.

With summarizing the properties analyzed above, we first give the correlation between the parameters and the performance of OFDM transceiver in Table 1, which is also presented in the simulation section. In the following, we develop an algorithm to obtain the precoding matrix **P** by selecting the frequencies $\boldsymbol{\omega}_b$ and $\boldsymbol{\omega}_a$, according to the summary in Table 1. The main spectrum envelope is decided in the initialization process step by step as

(a.) The value of $N_a$ and Case is selected according to the power spectrum envelope property.

(b.) $R$ is chosen according to sidelobe suppression performance and spectral efficiency.

(c.) $\alpha_0$ is the given condition number according to the BER quality constraint.

If Case A is required, we fix $\boldsymbol{\omega}_a$ far from the mainlobe and adjust $\boldsymbol{\omega}_b$. The process to solve (20) is given in the *algorithm. 1*.

---
**Algorithm. 1:**

**Initialization:** $\alpha_0$, $R$, $N_a$, iteration increment $\Delta\omega$, $\boldsymbol{\omega}_b = \boldsymbol{\omega}_{b_0}$, $\boldsymbol{\omega}_a = \boldsymbol{\omega}_{a_0}$, $\alpha = 0$, $i = 0$.

1. $i : i = i+1$. *i-th* iteration.
2. Design $\mathbf{P}_i$ by Section 3 and compute the condition number $\alpha = \mathrm{Con}(\mathbf{P}_i)$.
3. if $(\alpha \leq \alpha_0)$
    $\boldsymbol{\omega}_b := \boldsymbol{\omega}_b + \Delta\omega$, go back to Step 1.
   else
    Stop.

**Output:** $\mathbf{P} = \mathbf{P}_{i-1}$.

---

where $\boldsymbol{\omega}_{b_0}$ is the initializing frequency points close to the mainlobe, and $\boldsymbol{\omega}_{a_0}$ is the fixed frequencies far from the mainlobe. If the number of reserved carriers of $R$ is large, a little adjusting of $\boldsymbol{\omega}_b$ will lead to

Submitted

large change to $\alpha$. Therefore, the iteration increment $\Delta\omega$ also should be small, and the distance between the frequency points in $\boldsymbol{\omega}_b$ should not be too small. In practice, we select $R \leq 4$.

If Case B is required, we fix $\boldsymbol{\omega}_b$ far from the mainlobe and adjust $\boldsymbol{\omega}_a := \boldsymbol{\omega}_a + \Delta\omega$ in the Step 3. The distance between the frequency points in $\boldsymbol{\omega}_b$ also should not be too small or too far from the mainlobe. Otherwise, the matrix $\mathbf{C}_{op}$ may be close to singular or badly scaled. That may lead to that the results may be inaccurate in (13) by SVD in practice.

## 5. Simulation Results and Discussions

In this section, some numerical results are presented to demonstrate the performance of the proposed method. An instance in LTE is selected that the subcarrier spacing is 15 kHz. The number of the subcarriers is $N = 300$ in 5 MHz bandwidth [4]. The frequency axis is normalized to the spacing $2\pi/T$. The index of data subcarriers is from -150 to 150, while the direct current carrier $\omega = 0$ is not employed. To illustrate the OOB power reduction effect, the PSD is obtained by computing the power of DFT coefficients of time-domain OFDM signal over a time span and averaging over thousands of symbols. The frequency-domain oversampling rate is eight and the QPSK modulation is employed. The simulated are mainly ZP-OFDM system, unless noted otherwise.

*A. Sidelobe Suppression Performance*

In the first experiment, the OOB reduction performance comparison is presented using different precoding technologies. The number of reserved subcarriers is selected as $R = 2$. The SSOP [19], the orthogonal projection (OP) [16], the optimal orthogonal precoding (OOP) [21] and CC [8] methods are selected as a comparison. The zero-forcing frequencies are fixed at $\boldsymbol{\omega} = \pm 180$ for OP and SSOP method. The optimized region for OOP and CC scheme is also selected nearby the frequencies $\boldsymbol{\omega} = \pm 180$. $N_a = 2$ for double-side suppression to keep the spectrum symmetric in the proposed JSP method. The spectral

Submitted

coding rate is 300/302, $\omega_a = \pm 4000$ and $\omega_b = \pm 180$. The condition number of the precoding matrix is $\alpha = 3.2279$.

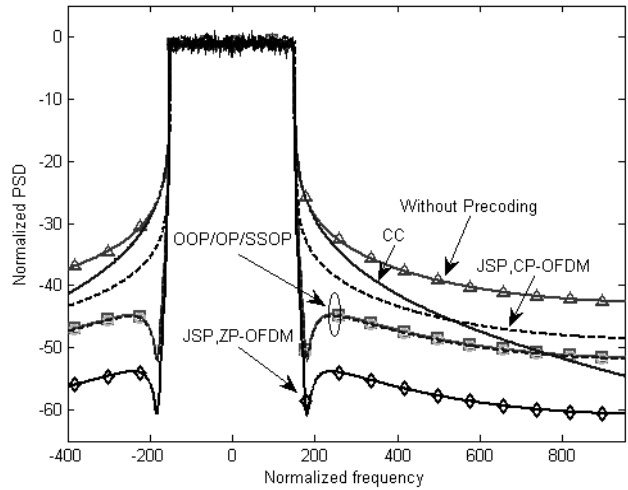

*Fig. 4* *PSD of the OFDM signal with different precoding techniques.*

As the PSD curves illustrated in Fig. 4, all the transmitted signals are ZP-OFDM, besides CP-OFDM signal is presented by the proposed JSP method. It is obviously that the OOB power using the JSP method is lower than other approaches. The OOP/OP/SSOP methods have similar suppression performance. The CC, which is not a spectral precoding method, achieves less OOB power reduction. In addition, the results show the performance degradation of sidelobe suppression in CP systems.

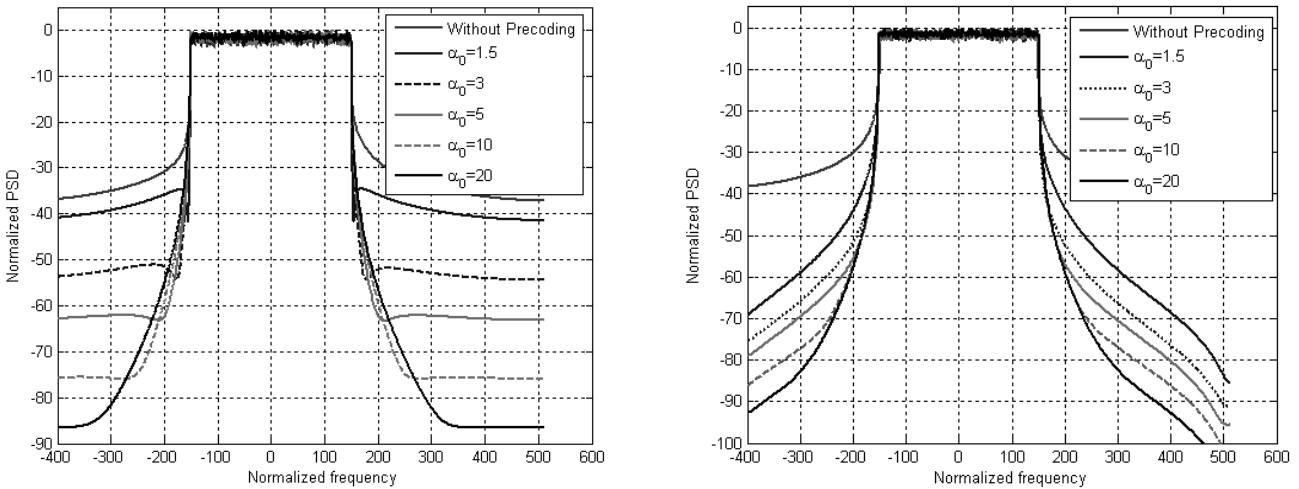

*Fig. 5* *Quicker power reduction on the edge of the mainlobe (Case A) with different condition number constraint.*

*Fig. 6 Lower power leakage far from the mainlobe (Case B) with different condition number constraint.*

In Fig. 5, the precoding matrix is designed under a condition number constraint of $\alpha_0 = \{1.5, 3, 5, 10, 20\}$ respectively. The PSD curves of Case A is presented with $N_a = 2$. $\boldsymbol{\omega}_a$ is fixed far from the mainlobe and $\boldsymbol{\omega}_a = \pm 4000$, $R = 2$ and $\Delta\omega = 0.25$. Obviously, the suppression performance is improved by relaxing the condition number constraint while the distance between $\boldsymbol{\omega}_b$ and mainlobe increasing. In Fig. 6, the Case B is presented with $N_a = 2$. We fix $\boldsymbol{\omega}_b$ far from the mainlobe and $\boldsymbol{\omega}_b = \pm 4000$, and the other parameters are same with in Fig. 5. The suppression performance is also improved by relaxing $\alpha_0$. The OOB power in the region far from the mainlobe is lower than Case A. However, the power on the edge of mainlobe decreases slowly.

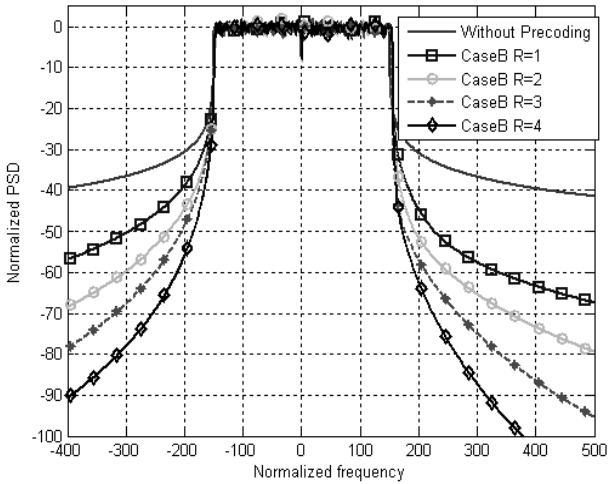 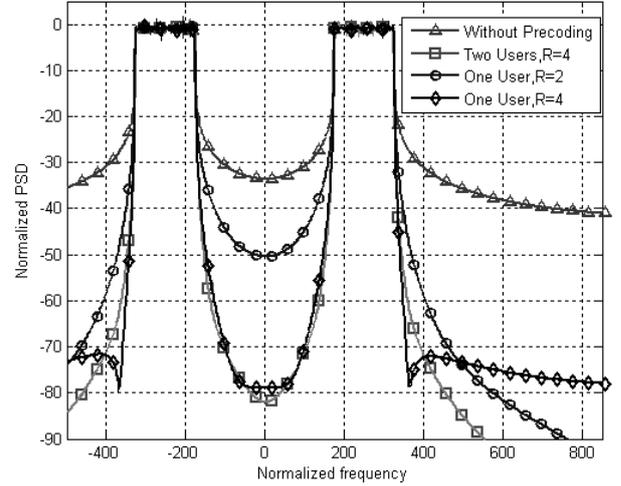

*Fig. 7 Single-side suppression with different number of revered carriers, $N_a=1$.*

*Fig. 8 PSD of Multi-band OFDM signal.*

In Fig. 7, the Case B is selected, $N_a = 1$ and $\alpha_0 = 1.5$. The right-side suppression is presented by choosing $\boldsymbol{\omega}_a$ and $\boldsymbol{\omega}_b$ in the right side of the mainlobe. The distance between two adjacent frequencies in $\boldsymbol{\omega}_a$ is five when $R > 1$. As the symmetry presented in (5), (9) and (14), the difference between adopting the OOB frequency $\boldsymbol{\omega}$ and $-\boldsymbol{\omega}$ to design the precoding matrix is only in the second step of SVD. Therefore, the power on the left side OOB region is also reduced, while the power emission is higher than on the right side. The suppression performance is improved by increasing the number of revered carriers. Although the simulation is not illustrated, this property is also presented in the case of $N_a = 2$ by adjusting $R$.

In cognitive radio system, the secondary users access the spectrum hole dynamically. In addition, carrier aggregation (CA) [31] is a critical technology in LTE system, which enable a user to employ non-contiguous subbands. Therefore, in Fig. 8, the PSD of two non-contiguous subbands occupied by one user or two is presented. The condition number is $\alpha_0 = 10$ for all the cases. The number of the carriers of each subband is 150. The design of the precoding matrix for two users is independent, while the data transmitted by one user through two subbands is dependent. The results indicate that the proposed scheme is also suitable for multi-users transmitting through non-contiguous subbands.

## B. BER performance

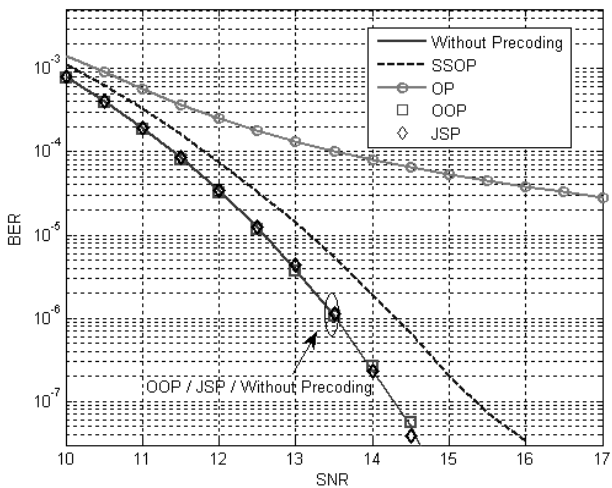 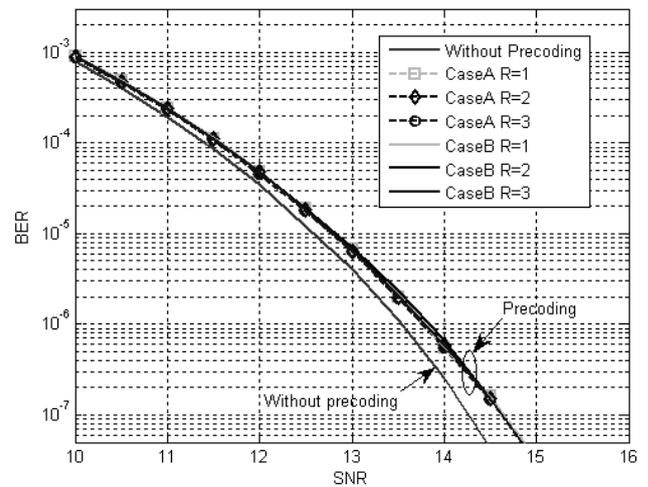

*Fig. 9* BER performance with different spectral precoding techniques through AWGN channel.

*Fig. 10* BER performance with same condition number constraint through AWGN channel, $N_a$=1.

A comparison of the BER performance using spectral precoding techniques is shown in Fig.9. The condition number constraint is $\alpha_0 = 2$ for the JSP. The optimized region or zero-forcing frequencies for OOP, SSOP and OP method is same with the JSP method. As illustrated in Fig. 4, the sidelobe suppression performance by OOP, SSOP and OP method is similar. The difference is that OOP scheme introduces no quality loss in the receiver and the OP method leads to large error as presented in Fig. 9. The SSOP method improves the receive quality by adopting a reserved subcarrier to decode the distorted signal. The BER quality loss by the JSP method is not notable.

In Fig. 10, the condition number constraint is fixed as $\alpha_0 = 10$, as well as single-side suppression is selected as $N_a = 1$. The results illustrated that the receive quality is approximate identical when the condition number and $N_a$ are fixed, although both of the number of reserved carriers $R$ and the Case are different. This property indicates that if $N_a$ has been selected according to single-side suppression or double-side suppression, the BER performance constraint can be converted to the condition number constraint in practice.

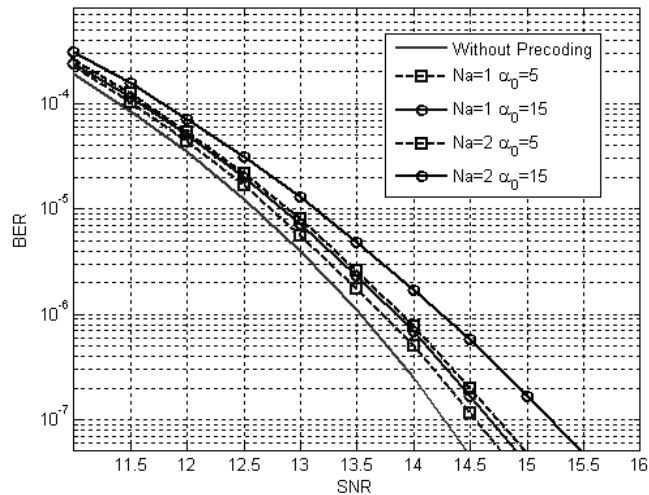

*Fig. 11* BER performance with difference condition number constraint and $N_a$ through AWGN channel, R=2.

In Fig. 11, the results demonstrate that the BER performance of $N_a = 1$ is better than $N_a = 2$ under same constraint. This is why we select the value of $N_a$ as small as possible to maintain the receive quality.

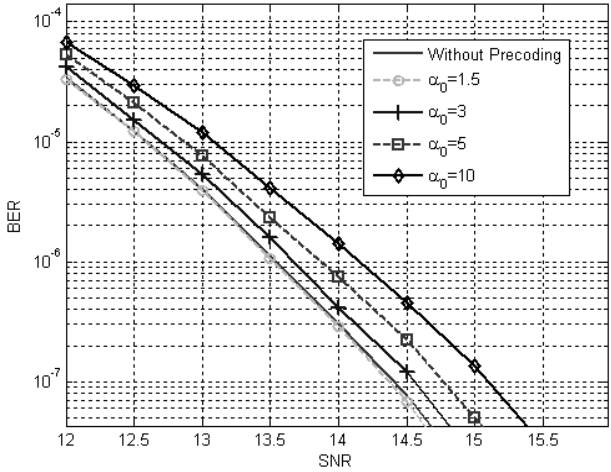 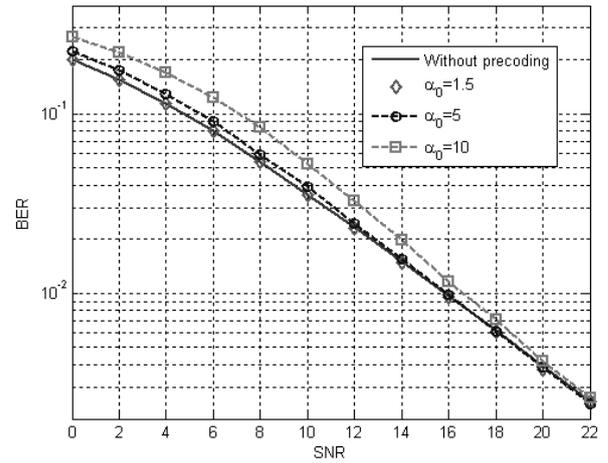

**Fig. 12** *BER performance with difference condition number constraint through AWGN channel.*

**Fig. 13** *BER performance with difference condition number constraint through 3GPP EPA fading channel.*

In Fig. 12, the receive quality is presented with $N_a = 2$. The number of revered carriers is $R = 2$. The results show that the BER performance decreases by relaxing $\alpha_0$, but the suppression performance is improved shown in Fig. 5-6. For example, the power in the OOB region can be reduced by nearly 40dB when $\alpha_0 = 10$, while the SNR loss is less than 1dB at BER=$10^{-7}$. When the condition number of precoding matrix is small, such as $\alpha_0 = 1.5$ in Fig. 12, the BER performance is slightly better than that of without precoding. Because the number of the modulated carriers of each symbol is extended from $N$ to $M$, the receive quality is improved by frequency diversity. However, this property is not notable when the value of $\alpha_0$ is large.

In Fig. 13, the parameters are set same with in Fig.12. Comparing to without precoding, the BER performance through the fading channel is similar to AWGN channel, when SNR is small. The receive quality decreases by relaxing $\alpha_0$. But with SNR increasing, the BER line converges to without precoding.

## 6. Conclusions

In this paper, a joint spectral precoding (JSP) method is proposed to reduce the OOB power emission in OFDM-based system, as well as an iteration algorithm is given to obtain the precoding matrix. We also convert the BER performance constraint to the condition number constraint of the precoding matrix. As summarized in Table I and presented in the simulations, the proposed method has an appreciate balance between suppression performance and spectral efficiency, under a receive quality constrain. We also can obtain the desirable spectrum envelope by parameters configuration. Therefore, this method is flexible for spectrum shaping in both conventional OFDM systems and non-continuous multi-carriers cognitive radio networks.

## 7. References

1. Xing, Y.P., Chandramouli, R., Mangold, S., and Shankar, N.S., 'Dynamic Spectrum Access in Open Spectrum Wireless Networks', *IEEE Journal on Selected Areas in Communications*, 2006, 24, (3), pp. 626-637.
2. Zhao, Q. and Sadler, B.M., 'A Survey of Dynamic Spectrum Access', *IEEE Signal Processing Magazine*, 2007, 24, (3), pp. 79-89.
3. Mahmoud, H.A., Yucek, T., and Arslan, H., 'Ofdm for Cognitive Radio: Merits and Challenges', *IEEE Wireless Communications*, 2009, 16, (2), pp. 6-14.
4. Dahlman, E., Parkvall, S., and Skold, J., *4g: Lte/Lte-Advanced for Mobile Broadband*, (Academic Press, 2013)
5. Faulkner, M., 'The Effect of Filtering on the Performance of Ofdm Systems', *IEEE Transactions on Vehicular Technology*, 2000, 49, (5), pp. 1877-1884.
6. Lin, Y.P. and Phoong, S.M., 'Window Designs for Dft-Based Multicarrier Systems', *IEEE Transactions on Signal Processing*, 2005, 53, (3), pp. 1015-1024.
7. Schmidt, J.F., Costas-Sanz, S., and Lopez-Valcarce, R., 'Choose Your Subcarriers Wisely: Active Interference Cancellation for Cognitive Ofdm', *IEEE Journal on Emerging and Selected Topics in Circuits and Systems*, 2013, 3, (4), pp. 615-625.
8. Brandes, S., Cosovic, I., and Schnell, M., 'Reduction of out-of-Band Radiation in Ofdm Systems by Insertion of Cancellation Carriers', *IEEE Communications Letters*, 2006, 10, (6), pp. 420-422.
9. Cosovic, I., Brandes, S., and Schnell, M., 'Subcarrier Weighting: A Method for Sidelobe Suppression in Ofdm Systems', *IEEE Communications Letters*, 2006, 10, (6), pp. 444-446.
10. Cosovic, I. and Mazzoni, T., 'Suppression of Sidelobes in Ofdm Systems by Multiple-Choice Sequences', *European Transactions on Telecommunications*, 2006, 17, (6), pp. 623-630.
11. Li, D., Dai, X.H., and Zhang, H., 'Sidelobe Suppression in Nc-Ofdm Systems Using Constellation Adjustment', *IEEE Communications Letters*, 2009, 13, (5), pp. 327-329.
12. Liu, S., Li, Y., Zhang, H., and Liu, Y., 'Constellation Expansion-Based Sidelobe Suppression for Cognitive Radio Systems', *Communications, IET*, 2013, 7, (18), pp. 2133-2140.

13. Mahmoud, H.A. and Arslan, H., 'Sidelobe Suppression in Ofdm-Based Spectrum Sharing Systems Using Adaptive Symbol Transition', *IEEE Communications Letters*, 2008, 12, (2), pp. 133-135.
14. Xu, R., Chen, M., Zhang, J., Wu, B., and Wang, H., 'Spectrum Sidelobe Suppression for Discrete Fourier Transformation-Based Orthogonal Frequency Division Multiplexing Using Adjacent Subcarriers Correlative Coding', *IET Communications*, 2012, 6, (11), pp. 1374-1381.
15. Huang., X., Zhang., J.A., and Guo, Y.J., 'Out-of-Band Emission Reduction and a Unified Framework for Precoded Ofdm', *IEEE Communications Magazine*, 2015, 53, (6), pp. 151-159.
16. van de Beek, J. and Berggren, F., 'N-Continuous Ofdm', *IEEE Communications Letters*, 2009, 13, (1), pp. 1-3.
17. Zheng, Y.M., Zhong, J., Zhao, M.J., and Cai, Y.L., 'A Precoding Scheme for N-Continuous Ofdm', *IEEE Communications Letters*, 2012, 16, (12), pp. 1937-1940.
18. van de Beek, J., 'Sculpting the Multicarrier Spectrum: A Novel Projection Precoder', *IEEE Communications Letters*, 2009, 13, (12), pp. 881-883.
19. Zhang, J.A., Huang, X.J., Cantoni, A., and Guo, Y.J., 'Sidelobe Suppression with Orthogonal Projection for Multicarrier Systems', *IEEE Transactions on Communications*, 2012, 60, (2), pp. 589-599.
20. Zhou, X.W., Li, G.Y., and Sun, G.L., 'Multiuser Spectral Precoding for Ofdm-Based Cognitive Radio Systems', *IEEE Journal on Selected Areas in Communications*, 2013, 31, (3), pp. 345-352.
21. Ma, M., Huang, X.J., Jiao, B.L., and Guo, Y.J., 'Optimal Orthogonal Precoding for Power Leakage Suppression in Dft-Based Systems', *IEEE Transactions on Communications*, 2011, 59, (3), pp. 844-853.
22. Tom, A., Sahin, A., and Arslan, H., 'Mask Compliant Precoder for Ofdm Spectrum Shaping', *IEEE Communications Letters*, 2013, 17, (3), pp. 447-450.
23. Muquet, B., Wang, Z.D., Giannakis, G.B., de Courville, M., and Duhamel, P., 'Cyclic Prefixing or Zero Padding for Wireless Multicarrier Transmissions?', *IEEE Transactions on Communications*, 2002, 50, (12), pp. 2136-2148.
24. Lu, H., Nikookar, H., and Chen, H., 'On the Potential of Zp-Ofdm for Cognitive Radio', *Proc. WPMC'09*, 2009, pp. 7-10.
25. Van Waterschoot, T., Le Nir, V., Duplicy, J., and Moonen, M., 'Analytical Expressions for the Power Spectral Density of Cp-Ofdm and Zp-Ofdm Signals', *Signal Processing Letters, IEEE*, 2010, 17, (4), pp. 371-374.
26. Wu, T.W. and Chung, C.D., 'Spectrally Precoded Dft-Based Ofdm and Ofdma with Oversampling', *IEEE Transactions on Vehicular Technology*, 2014, 63, (6), pp. 2769-2783.
27. Meyer, C.D., *Matrix Analysis and Applied Linear Algebra*, (SIAM, 2000)
28. Tong, J., Guo, Q.H., Tong, S., Xi, j.T., and Yu, Y.G., 'Condition Number-Constrained Matrix Approximation with Applications to Signal Estimation in Communication Systems', *IEEE Communications Letters*, 2014, 21, (8), pp. 990-993.
29. 3GPP TS 36.141: 'E-Utra Base Station (Bs) Conformance Testing', 2015

Submitted

30. Aubry, A., De Maio, A., Pallotta, L., and Farina, A., 'Maximum Likelihood Estimation of a Structured Covariance Matrix with a Condition Number Constraint', *Signal Processing, IEEE Transactions on*, 2012, 60, (6), pp. 3004-3021.
31. Yuan, G.X., Zhang, X., Wang, W.B., and Yang, Y., 'Carrier Aggregation for Lte-Advanced Mobile Communication Systems', *IEEE Communications Magazine*, 2010, 48, (2), pp. 88-93.